# From Neuronal Packets to Thoughtseeds: A Hierarchical Model of Embodied Cognition in the Global Workspace

**Prakash Chandra Kavi**[1*]

**Gorka Zamora Lopez**[1]

**Daniel Ari Friedman**[2]

[1]**Center for Brain and Cognition, Universitat Pompeu Fabra, Barcelona, Spain**
[2]**Active Inference Institute, Davis, California, USA**

*Corresponding Author

**Keywords:** Global Workspace Theory[1], evolution[2], embodied cognition[3], Markov Blanket[4], thoughtseed[5] , Free Energy Principle[6] , neuronal packet[7]

## Abstract

The emergence of cognition requires a framework that bridges evolutionary principles with neurocomputational mechanisms. This paper introduces the novel "thoughtseed" framework, proposing that cognition arises from the dynamic interaction of self-organizing units of embodied knowledge called "thoughtseeds" within the Global Workspace of consciousness. Leveraging foundational concepts from evolutionary theory, neuronal packets, and free energy principle, we propose a hierarchical model of cognitive states, comprising Neuronal Packet Domains (NPDs), Knowledge Domains (KDs), the thoughtseed network, and meta-cognition. This hierarchical interplay, mediated by nested Markov blankets and reciprocal message passing, facilitates the emergence of thoughtseeds as coherent patterns of activity that guide perception, action, and learning. Thoughtseeds, posited as fundamental units of thought, compete for dominance within the Global Workspace, with the dominant thoughtseed shaping conscious experience and guiding behavior. We present a mathematical framework grounded in active inference and dynamical systems theory to model thoughtseed dynamics and their contribution to the unitary nature of consciousness. The thoughtseed framework offers a promising step towards a novel, biologically-grounded model for understanding the organizing principles and emergence of embodied cognition, offering a unified account of cognitive phenomena, with potential applications in understanding consciousness, attention, and decision-making.

# 1. Introduction

## Embodied Cognition: An Evolutionary and Free Energy Perspective

Survival depends not only on adaptation but also on the ability of living systems to actively shape their environments (Futuyma, 2017). Evolutionary mechanisms such as the Baldwin effect and natural selection explain how learning and environmental pressures influence evolutionary processes (Baldwin, 1896; Darwin, 1859). Beyond genetic evolution, modern perspectives recognize the role of epigenetic, behavioral, and cultural inheritance in shaping living systems and their environments (Jablonka & Lamb, 2005; Boyd & Richerson, 1988). These interactions underscore how living systems dynamically regulate their relationship with the environment, not just responding to it, but influencing it through actions and perceptions (Laland et al., 2000; Odling-Smee et al., 2003). This dynamic, adaptive relationship is central to the concept of embodied cognition, where understanding emerges through the interaction of the brain, body, and environment (Varela, Thompson & Rosch, 1991; Noë, 2004).

In line with this view, autopoiesis describes the self-organizing nature of living systems (Maturana & Varela, 1980), allowing them to maintain a non-equilibrium steady state (NESS) by continuously exchanging energy and matter with their surroundings (Friston & Ao, 2012 ). These systems resist the natural tendency towards disorder (entropy), exhibit emergent properties (Schrödinger, 1944; Nicolis & Prigogine, 1977) and actively regulate internal states to ensure persistence. Building on the idea that effective regulation requires an internal model (Conant & Ashby, 1970), the Free Energy Principle (FEP) provides a unifying framework to explain how living systems adapt by minimizing surprise or variational free energy (Friston, 2010). Through a process called active inference, living systems continuously refine their internal models and take actions that align their predictions with incoming sensory data (Friston et al., 2010; Parr et al., 2022). The Markov blanket concept, central to the FEP, separates internal states from the external world, enabling computational autonomy and *conditional independence* while mediating interactions with their environment (Friston, 2010; Parr et al., 2018).

FEP's scale-free modeling approach integrates both evolutionary and cognitive dynamics, offering insights into how adaptive systems emerge and evolve (Friston et al., 2021, 2024). Furthermore, the Hierarchically Mechanistic Mind (HMM) hypothesis views the brain as an adaptive control system that minimizes free energy through recursive interactions between neurocognitive processes, emerging as a result of evolutionary pressures and self-organization (Badcock et al., 2019).

## Neuronal Representations

Efforts to understand the neural basis of cognition have explored representational units beyond individual neurons, such as *cognits* (Fodor, 1983), *engrams* (Josselyn & Tonegawa, 2020) and *cell assemblies* (Hebb, 1949). Furthermore, recent research suggests that cognitive processes may operate in a discrete manner, with distinct packets of information or 'cognitive atoms' forming the building blocks of thought and consciousness (Dehaene, 2014). These discrete units of cognitive processing compete for access to a global workspace, where they can be broadcast across the brain to generate a unified and coherent conscious experience (Dehaene et al., 2011, 2003).

The Neuronal Packet Hypothesis (NPH) (Yufik & Friston, 2016; Palacios et al., 2020; Ramstead et al., 2021) posits that *neuronal packets* (NPs), self-organizing ensembles of neurons, serve as the *fundamental units of neuronal representation* in the brain (Yufik, 2019). Superordinate ensembles (SEs) emerge from the coordinated activity of multiple NPs, enabling the representation of complex and abstract concepts. Through their interactions and emitted signals, NPs allow the system to infer probabilistic beliefs about external states, continuously updating these beliefs to minimize free energy (Ramstead et al., 2021).

Each NP and SE possesses its own Markov blanket, which defines its boundaries and interactions. The Markov blankets of lower-level NPs are nested within those of higher-level SEs, integrating information through a shared generative model (Palacios et al., 2020). This hierarchical structure facilitates the brain's ability to represent knowledge across multiple scales, from sensory details to abstract categories. Nested SEs interact via reciprocal message passing, where top-down predictions from higher-level SEs modulate lower-level SEs and NPs, while bottom-up sensory evidence or prediction errors update higher-level representations. This continuous exchange of information supports adaptive behavior in response to a changing environment (Kirchhoff et al., 2018; Palacios et al., 2020; Friston et al., 2021). The FEP also has implications for the nature of representation itself,suggesting that the content of neural representations lies in their role in guiding action and minimizing surprise (Ramstead et al., 2020, 2023).

NPH lays a robust foundation for exploring several key questions: How do evolutionary pressures shape the emergence and organization of neuronal packets and their hierarchical ensembles? How do individual experiences interact with inherited predispositions to refine and adapt these representations? How do these representational units collaborate within a network to generate complex cognition and adaptive behavior?

# 2. Architectural Foundations of Embodied Cognition

| Concept | Symbol | Explanation |
| --- | --- | --- |
| Neuronal Packet (NP) | $\nu$ | The fundamental unit of neuronal representation, a self-organizing ensemble of neurons that encodes a specific feature or aspect of the world. |
| Core Attractor | $\psi_{\nu-c}$ | The most probable and stable pattern of neural activity within a manifested NP, embodying its core functionality. |
| Subordinate Attractor | $\psi_{\nu-s_i}$ | Less dominant patterns of neural activity within an NP that may become active under specific conditions or in response to novel stimuli, offering flexibility and adaptability. |
| Encapsulated Knowledge Structure | $\omega_\nu$ | The structured knowledge content within a NP's Markov Blanket, shaping its activity and contribution to KDs. |
| Superordinate Ensemble (SE) | $\mathcal{E}$ | A higher-order organization emerging from the coordinated activity of multiple NPs, or even KDs, enabling the representation of more complex and abstract concepts. |
| Neuronal Packet Domain (NPD) | $N$ | A functional unit within the brain, comprised of interconnected SEs, specialized for specific cognitive processes or tasks. |
| Knowledge Domain (KD) | $K$ | A large-scale, organized structure within the brain's internal model, representing interconnected networks of concepts, categories, and relationships that constitute a specific area of knowledge or expertise. |
| Thoughtseed | $\mathcal{T}$ | A higher-order construct with agency, emerging from the coordinated activity of SEs across different KDs. It represents a unified and meaningful representation of a concept, idea, or percept and guides perception, action, and decision-making. |
| Thoughtseed Activation Level | $\alpha_{\mathcal{T}}$ | A measure of a thoughtseed's prominence in the current cognitive landscape, calculated by weighing the probabilities that the brain's state aligns with the thoughtseed's dominant or subordinate attractor states. |
| Activation Threshold | $\Theta_{activation}$ | A global parameter, influenced by the consciousness state and arousal levels that determines the minimum activation level a thoughtseed must cross to enter the active thoughtseed pool. |
| Active Thoughtseed Pool | $\mathcal{P}_{active}$ | The set of active thoughtseeds whose activation levels surpass the activation threshold at a given time. |
| Dominant Thoughtseed | $\mathcal{T}^*$ | The thoughtseed within the active thoughtseed pool that has the highest activation level and is primarily shaping the content of consciousness at a given moment. |
| Thoughtseed Network | $TN$ | The collection of interconnected thoughtseeds within the brain's internal states, hypothesized to encode a generative model of the environment. |

**Table 1. Key Notations of Thoughtseed Framework**

## The Thoughtseed Hypothesis

A fundamental challenge in elucidating the neural basis of cognition lies in the inherent opacity of the Markov blanket, that can be referred to as a "black box" problem. While this statistical construct offers a robust framework for modeling the exchange of information between an agent and its environment, it inherently obscures the internal generative model responsible for behavior (Friston 2010, 2020).

To address these challenges, we propose the "thoughtseed" framework, which builds upon the concept of neuronal packets (NPs) and the neuronal packet hypothesis (NPH). This framework integrates insights from the Free Energy Principle (FEP) (Friston 2010, 2021), evolutionary theory, and Global Workspace Theory (GWT) (Baars, 1997; Dehaene & Naccache, 2001).

The thoughtseed hypothesis posits that cognition arises from the dynamic interaction of self-organizing units of embodied knowledge termed "thoughtseeds." These thoughtseeds, conceptualized as emergent agents with their own Markov blankets, are shaped by a hierarchy of evolutionary priors that encompass both inherited predispositions and learned experiences. The framework proposes a four-level hierarchical/heterarchical model of the cognitive agent's internal states: Neuronal Packet Domains (NPDs), Knowledge Domains (KDs), the Thoughtseed Network (TN), and meta-cognition.

Each level in this model consists of nested Markov blankets, encoding information about the world at varying levels of abstraction, with deeper levels representing more domain-general and abstract knowledge (Fields et al., 2022; Ramstead et al., 2023). The dynamic interplay within this hierarchy facilitates the emergence of thoughtseeds as coherent patterns of activity that guide perception, action, and learning. The framework emphasizes the embodied nature of cognition, where the living system's interactions with its *Umwelt* (von Uexküll, 1934) and the knowledge encoded in the KDs shape the Thoughtseed Network.

In the following sections, we will elaborate on the architectural foundations of the thoughtseed framework, exploring the details of NPDs, KDs, and their dynamic interplay in the emergence of thoughtseeds and the shaping of conscious experience through nested Markov blankets.

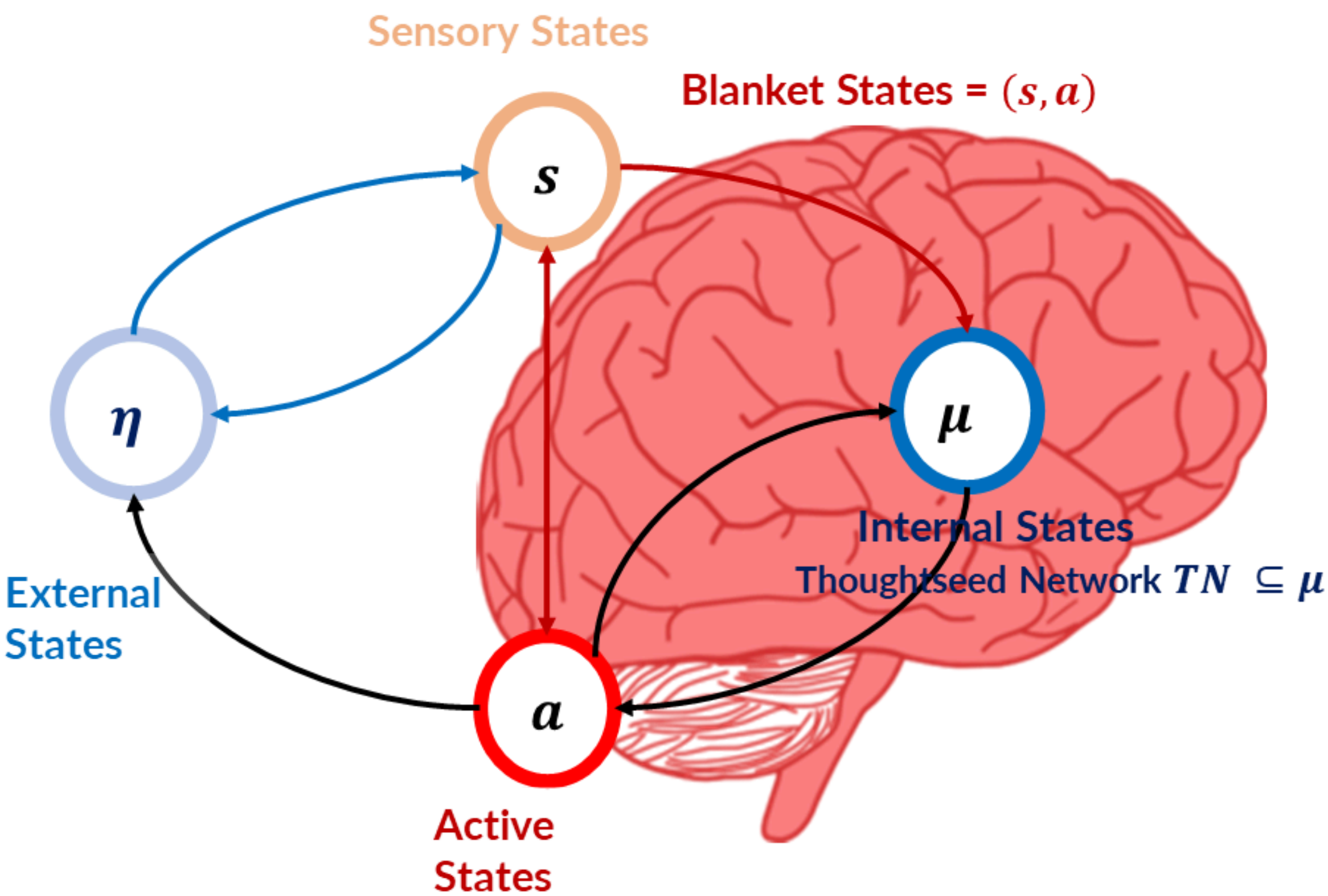

**Figure 1. Markov Blanket of an Agent.** Adapted from Friston et al., 2023, this diagram illustrates the partitioning of an agent's states into internal states (μ), encompassing the Thoughtseed Network (TN), and external states (η). The Markov blanket, comprising sensory (s) and active states (a), mediates the interaction between the internal and external states. The internal states, housing the TN, can only influence active states, while external states can only influence sensory states. This separation allows the TN to operate with a degree of autonomy, generating predictions and selecting actions based on its internal model of the world.

## Neuronal Packet Domains

The brain's sparse architecture, characterized by a limited number of active neurons and low connection density, supports localized functional units for specific cognitive tasks (Bullmore & Sporns, 2009; Sporns & Betzel, 2016). NPDs are hypothesized to emerge through self-organizing interconnected NPs. Their hierarchical organization, via nested Markov blankets, reflects "ascending scales of canonical microcircuits" (Hipólito et al., 2021; Douglas & Martin, 1991; Mountcastle, 1997), enabling complex computations from simpler units. This structure facilitates efficient information processing and adaptive behavior (Kirchhoff et al., 2018; Palacios et al., 2020; Friston et al., 2021).

NPD formation is guided by evolutionary priors that favor adaptive neural architectures suited for specific functions (Quartz & Sejnowski, 1997). For instance, the visual cortex has distinct NPDs for color, motion, and form (Zeki, 1993 ; DiCarlo et al., 2012), with these specializations evolutionarily conserved across mammalian species (Markov et al., 2014). Hebbian learning further shapes their specific organization, as elaborated in Section 2.4 and 8.2.

NPs compete for resources by minimizing variational free energy (VFE), resulting in efficient internal models. This competition, described as "neural Darwinism," produces specialized NPs, each representing different aspects of the world (Edelman, 1987). The Markov blanket structure supports this decentralized, modular organization, enhancing both robustness and adaptability.

NPs dynamically interact to form superordinate ensembles (SEs) within NPDs, representing higher-order concepts and allowing for greater complexity in cognitive representations. The hierarchical structure, where the Markov blankets of lower-level NPs are nested within those of higher-level SEs, enables the brain to encode knowledge across different scales, from sensory details to abstract categories (Riesenhuber & Poggio, 2000; Ramstead et al., 2021). SEs may emerge as stable entities when they accurately predict and explain sensory input (Friston et al., 2021). The nested SEs interact through reciprocal message passing, facilitating flexible and adaptive behavior in response to changing environments.

Furthermore, the formation of SEs within NPDs is influenced by the alignment of knowledge representations through vector rotations. When multiple NPs contribute to a shared generative model, their encapsulated knowledge representations undergo a process of alignment through vector rotations (Yufik & Friston, 2016). This rotation facilitates the integration of knowledge from different NPs by bringing their vectors closer together in the knowledge space. The rotation angle is determined based on the similarity between the generative models of the participating NPs. This alignment process is an integral part of the formation of SEs, ensuring that the constituent NPs have a coherent and integrated representation of the knowledge relevant to the shared generative model.

## Neuronal Packets (NPs): The Fundamental Units of Neuronal Representation

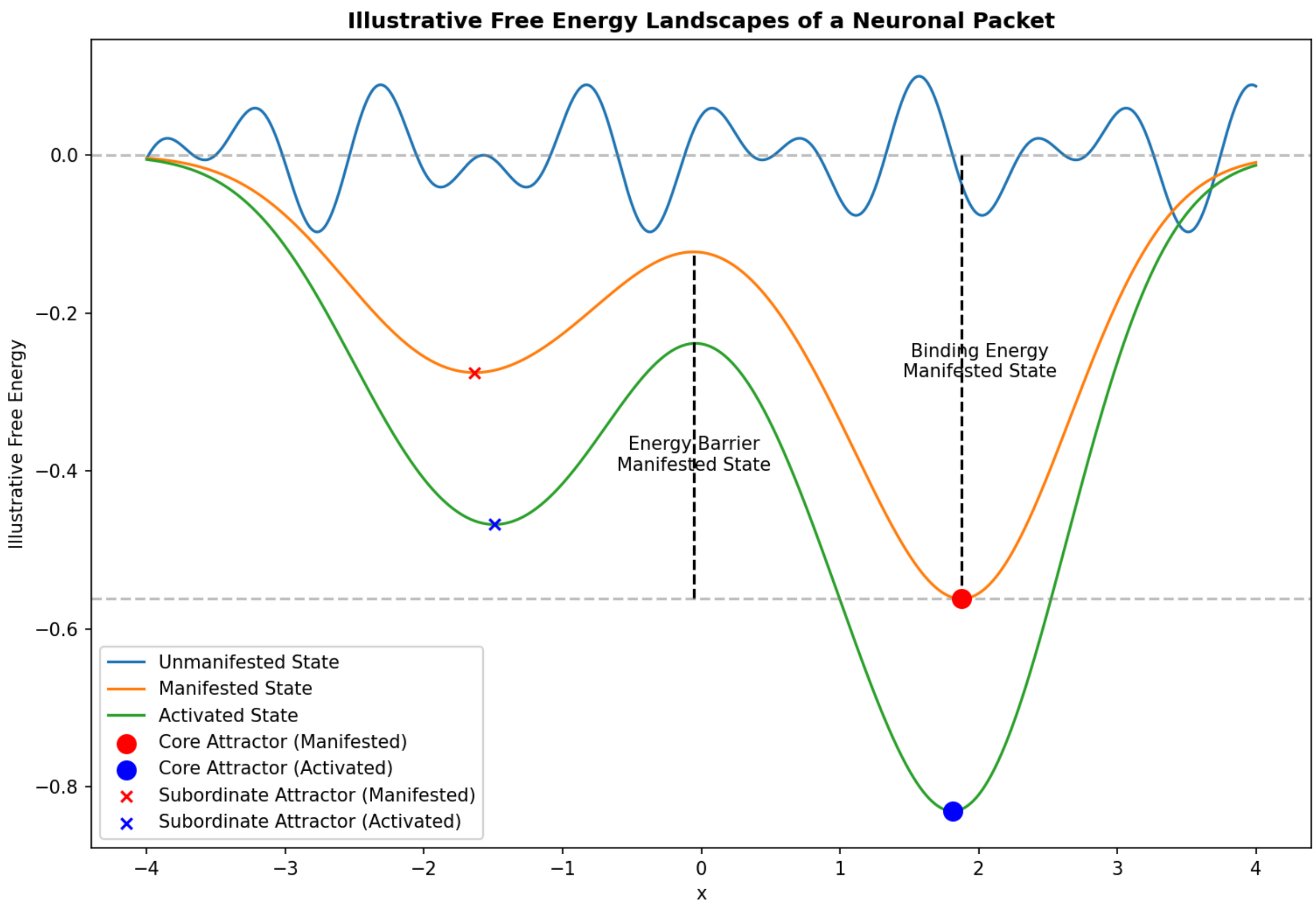

**Figure 2. Illustrative Free Energy Landscapes of a Neuronal Packet** The figure depicts the illustrative free energy landscapes associated with the three potential states of a neuronal packet (NP). The x-axis represents the hypothetical internal states of the NP, while the y-axis represents the free energy associated with each state. The blue curve depicts the **unmanifested state**, characterized by a relatively flat landscape with shallow local minima, indicating low stability and high susceptibility to change. Upon undergoing a **phase transition** from the unmanifested state to the **manifested state**, the NP could be represented by the red curve, featuring a deep local minimum (the **core attractor**) and a shallower local minimum (a **subordinate attractor**). The **energy barrier** associated with the manifested state is depicted by the vertical dashed line, highlighting the energy required to transition between the core attractor and the subordinate attractor. The **binding energy**, representing the overall stability of the manifested NP, is visualized as the vertical distance between the core attractor and the zero free energy level (horizontal dashed line). The green curve illustrates the activated state, where the core attractor is further deepened, reflecting heightened neural activity and a stronger focus on the NP's core representation.

Within each NPD, NPs serve as the fundamental units of neuronal representation (Ramstead et al., 2021; Yufik, 2016, 2019). An NP could exist in three states:

- **Unmanifested State:** Represents a potential configuration of neural activity with high prior probability under specific conditions, shaped by evolutionary priors. It can be viewed as a sparsely connected neural ensemble with low precision, corresponding to a shallow local minimum in the free energy landscape, indicating low stability and high potential for change.
- **Manifested State:** Emerges from the unmanifested state upon repeated exposure to relevant stimuli, leading to a **phase transition** (Yufik, 2019) and the formation of a Markov Blanket stabilized by an **energy barrier** (Yufik, 2019). This Markov blanket structure may enable local computation and autonomy within the NP, while maintaining informational boundaries. It is characterized by increased coherence of neural activity, resulting in a stable state with a **core attractor** representing the most probable and stable pattern of neural activity (Friston & Ao, 2012; Parr & Friston, 2018). This state embodies the NP's core functionality with high certainty and can be interpreted as the mode of the posterior distribution over the NP's internal states, given its Markov blanket (Friston et al., 2021). It corresponds to a deeper local minimum in the free energy landscape, reflecting high stability and low surprise (Kiebel et al., 2009). The depth of this global minimum could indicate the NP’s **binding energy**, reflecting the overall stability and resistance to change of the core representation (Yufik, 2019). **Alternative attractors** represent less dominant patterns of neural activity that may become active under specific conditions or in response to novel stimuli (Rabinovich et al., 2008). They can be viewed as shallower local minima in the free energy landscape, separated from the core attractor by **energy barriers**. The existence of alternative attractors, along with the energy barriers that separate them, allows for flexibility and adaptability in the NP's response to changing inputs.

- **Activated (or Spiking) State:** A transient state characterized by heightened neural activity within the manifested NP ensemble, triggered by specific inputs that resonate with the NP's internal model. The NP generates a response that may influence the living system's behavior or cognition. This response can be interpreted as a consequence of the NP's internal dynamics and its attempt to minimize free energy, rather than a deliberate or planned action. The active state can be seen as a temporary shift in the free energy landscape, where the core attractor becomes even more pronounced.

## Knowledge Domains (KDs)

Knowledge Domains (KDs) can be conceptualized as large-scale, organized structures within the brain's internal model, akin to knowledge graphs (Hogan et al., 2021). They serve as nested levels of "knowledge repositories" that provide the conceptual scaffolding for interpreting sensory information and generating the content of consciousness within the Global Workspace. KDs encompass not only sensory representations but also retrieved memories, beliefs, experiences, policies, emotions, and learned patterns.

KDs are hypothesized to exhibit both hierarchical and heterarchical structures, enabling flexible and context-dependent knowledge retrieval for adaptive behavior (Pessoa, 2014). The hierarchical aspect reflects the layered organization of knowledge, ranging from concrete to abstract (Friston, 2008; Hipólito et al., 2021; Zamora-López et a., 2011), while the heterarchical aspect captures the interconnectedness and cross-domain interactions among different KDs (Yufik, 2019).

NPDs function as specialized units that process specific information and project it into the Global Workspace. In contrast, KDs interpret and contextualize this raw data within the brain's neural architecture. The formation and adaptation of KDs are influenced by evolutionary priors, which provide a foundational structure for their development, as illustrated in Figure 3. Continuous learning throughout the lifespan further refines KDs (Spelke & Kinzler, 2007).

KDs possess an intrinsic affective dimension, reflecting the emotional valence and arousal associated with their content (Patisappu et al., 2024). This dimension is crucial for shaping subjective experiences and influencing behavior and decision-making processes (Sladky et al., 2021; Solms, 2021). However, KDs representing purely abstract concepts, such as mathematical or grammatical rules, may lack this affective component.

The dynamic and context-dependent binding process within KDs contributes to the emergence of coherent percepts in the Global Workspace. This process may involve the synchronization of neural activity across different NPDs (Singer, 1999; Palacios et al., 2019), the formation of superordinate ensembles encompassing multiple NPs, and the modulation of attentional mechanisms to prioritize relevant information. Enhanced firing rates in specific neurons may facilitate this binding across NPDs (Roelfsema, 2023). This process is further supported by reciprocal message passing and attentional selection, reflecting "circular causality" within the FEP, where different levels of the hierarchy mutually influence one another through prediction and error correction (Friston & Kiebel, 2009).

A contrived illustration of potential brain areas, having a greater influence on KDs might include:

- The hippocampus, serving as a central hub for a "memory" KD, integrating information to create episodic and semantic memories (Squire et al., 2015; Eichenbaum, 2017).
- The amygdala, associated with an "emotion" KD, linking sensory information with affective responses (LeDoux & Brown, 2017; Phelps & LeDoux, 2005).
- The prefrontal cortex, acting as a central hub for multiple KDs related to executive functions and goal-directed behavior (Miller & Cohen, 2001; Fuster, 2015).

The thoughtseed framework offers a biologically grounded model of cognition within the broader context of mortal computation (Ororbia & Friston, 2023), elucidating how cognitive processes emerge from the dynamic interplay of neuronal packets, knowledge domains, and higher-order structures. An illustrative example of the Visual Object Recognition KD is provided in Supplementary Section 9.

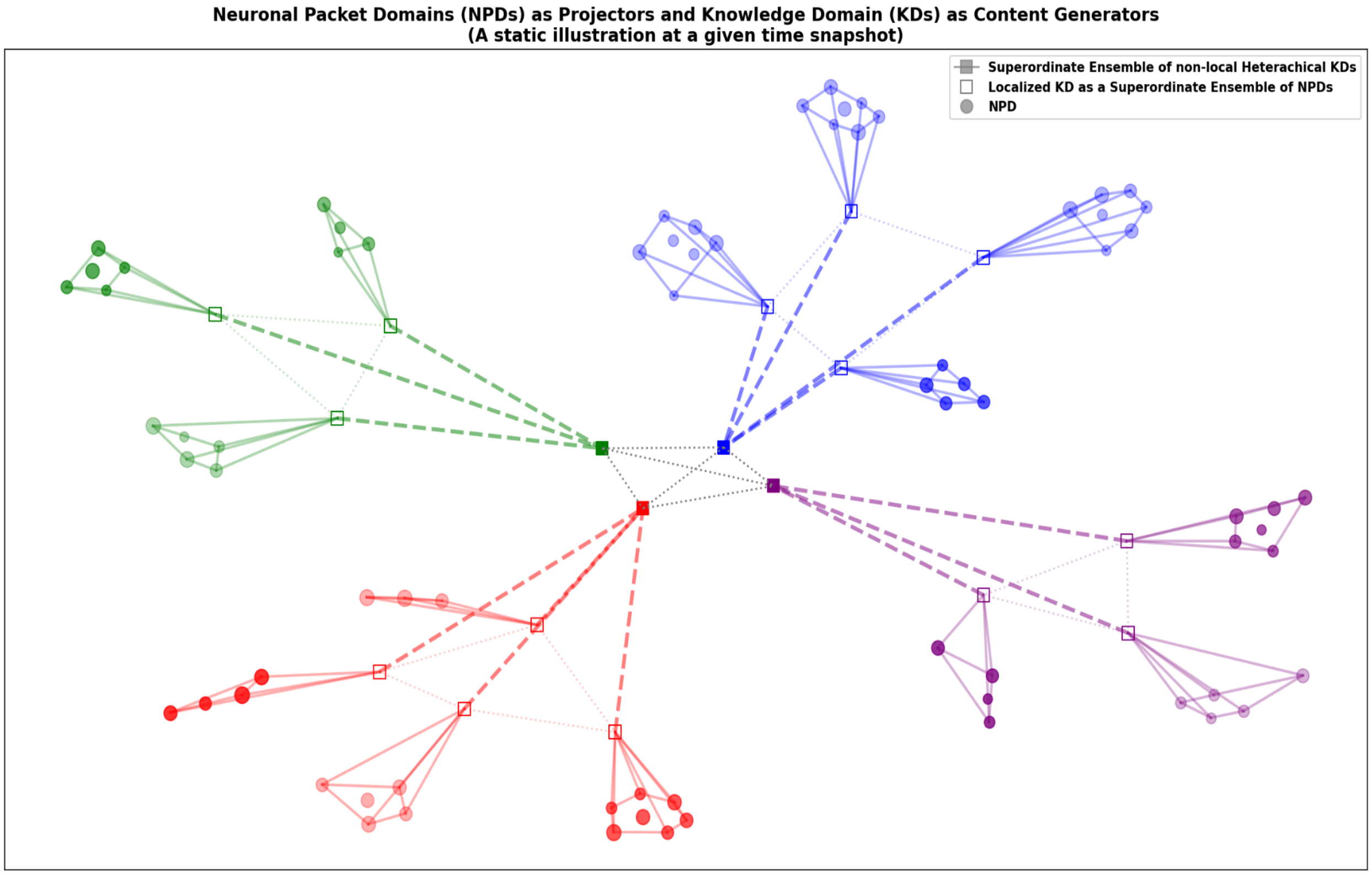

**Figure 3. Neuronal Packet Domains (NPDs)** and **Knowledge Domains (KDs).** This figure provides a static illustration of the relationship between NPDs and KDs at a given time snapshot, highlighting their roles in the thoughtseed framework. **KDs:** The four distinct colors represent different KDs, each specializing in a particular domain of knowledge or expertise. The unfilled squares symbolize *localized* KDs, formed as superordinate ensembles of NPDs, integrating information from specific functional units within the brain. The filled squares represent a superordinate ensemble of *non-local, heterarchical* KDs, suggesting the integration of knowledge across multiple domains. **NPDs:** The circles depict NPDs, the functional units of the brain responsible for processing specific types of information. They act as "projectors," providing raw sensory data or generating potential actions. **Hierarchical/Heterarchical Organization:** The arrangement of NPDs within or connected to KDs visually represents the hierarchical nature of knowledge representation, where KDs integrate and interpret information from multiple NPDs to form meaningful and complex representations. The presence of a higher-order KD further suggests the potential for hierarchical/heterarchical organization of knowledge domains themselves. **Dynamic Interplay:** While the figure provides a static snapshot, it implies the dynamic interplay between NPDs and KDs. The connections between them suggest the flow of information and influence, where NPDs provide raw data, and KDs interpret and contextualize this information, contributing to the formation of thoughtseeds. **Thoughtseed Formation and the Global Workspace:** Thoughtseeds emerge from the dynamic interplay between NPDs and KDs, drawing upon the knowledge and information represented within these domains. These thoughtseeds then compete for dominance within the Global Workspace, with the winning thoughtseed shaping the "content of conscious experience."

# 3. Thoughtseeds Network and Meta-cognition

## Formal Definition of a Thoughtseed

The thoughtseed framework posits that thoughtseeds are self-organizing, higher-order cognitive constructs that emerge from the coordinated activity of distributed neural networks, aligning with research on large-scale brain networks and the Global Workspace Theory of consciousness (GWT) (Mashour et al., 2020).

Thoughtseeds can be understood as transient units within this global workspace, representing temporary coalitions of neural activity that form, compete, and dissolve as the individual interacts with the world. These dynamic patterns of neuronal activity can be interpreted as metastable states within the brain's pullback attractor landscape (Friston, 2010; Friston & Ao, 2012; Friston, Sengupta, & Auletta, 2014).

The content and structure of thoughtseeds emerge from the knowledge and beliefs embedded within Knowledge Domains (KDs), reflecting the living system's evolutionary history and individual experiences. The dynamic interplay between thoughtseeds and KDs shapes cognition, behavior, and conscious experience.

**Key Characteristics of Thoughtseeds:**

- **Sub-Agents:** Thoughtseeds act as autonomous sub-agents within the cognitive system, engaging in active inference to generate predictions, influence actions, and update internal models based on sensory feedback (Seth, 2015). This agency allows them to explore the environment and develop affordances.

- **Pullback Attractor Dynamics:** When active, thoughtseeds function as pullback attractors (Friston & Ao, 2012; Haken, 1983), integrating information from multiple KDs to form coherent representations. This establishes a transient Markov blanket, maintaining autonomy and computational independence. Each thoughtseed is associated with a core attractor, representing its most probable and stable pattern of neural activity, and may also have subordinate attractors, offering flexibility and adaptability.
- **Goal-Directed Behavior:** Thoughtseeds are inherently goal-directed, driven by the imperative to minimize EFE (Parr & Friston, 2019). They anticipate action consequences and select those that minimize EFE. The interaction between thoughtseeds and KDs results in affordances—potential actions within the environment (Gibson, 1977). These affordances can be categorized as epistemic (opportunities for learning or exploration) and pragmatic (opportunities for goal fulfillment or exploitation) (Friston et al., 2015; Friston, 2022).
- **Thoughtseed States:** Thoughtseeds exist in the following states: unmanifested, manifested (inactive, active, dominant), and dissipated.
    - **Unmanifested:** The thoughtseed exists as a potential configuration within the neural network, latent within interconnected knowledge domains, primed by prior experiences and learning. It is not yet actively influencing cognition or behavior.
    - **Manifested:** The thoughtseed has emerged and is now part of the cognitive landscape, with the potential to influence thought and action. This state is further divided into the following sub-states:
        - **Inactive:** The thoughtseed is present but not currently contributing to conscious experience. It resides in the background, with its core attractor and associated dynamics intact, potentially primed for activation.
        - **Active:** The thoughtseed is in the "active thoughtseed pool" and contributes to the content of consciousness, competing with other active thoughtseeds for dominance.

- **Dominant:** The dominant thoughtseed is selected through an ongoing process of cumulative EFE minimization, where thoughtseeds within the active thoughtseeds pool compete in a winner-take-all dynamic. The thoughtseed that minimizes cumulative EFE over time gains dominance, enters the Global Workspace, and shapes the unitary conscious experience. This process ensures that the most relevant, predictive, and adaptive thoughtseed guides cognition and action.
    - **Dissipated:** The thoughtseed has lost its influence and its core attractor has become significantly weakened or dissolved. It no longer contributes to the active thoughtseed pool and its constituent elements may be reabsorbed into the broader neural network or contribute to the formation of new thoughtseeds.

**Formal Definition:** A "thoughtseed" is a higher-order, transient, Markov-blanketed unit with agency that emerges from the coordinated activity of neuronal ensembles across different knowledge domains (KDs). It is characterized by a core attractor and subordinate attractors that shape its content and dynamics. Thoughtseeds exhibit a repetitive behavior, revisiting the brain's cognitive landscape even when not explicitly required, contributing to phenomena such as mind-wandering. Through continuous competition with other thoughtseeds via EFE minimization, a dominant thoughtseed emerges, enters the Global Workspace, and contributes to the unitary conscious experience, enabling active exploration of the environment and the development of affordances.

An illustrative example of a "dog thoughtseed" is shown in Supplementary Section 9.

## Thoughtseeds and the Global Workspace

The Global Workspace Theory (GWT) proposes that consciousness arises from a "global workspace" – a central information exchange hub within the brain (Baars, 1997; Dehaene & Changeux, 2011). This workspace facilitates communication and integration among specialized, modular processing units. Within the thoughtseed framework, Knowledge Domains (KDs) can be conceptualized as these specialized modules, each encapsulating a specific domain of knowledge or expertise. These KDs operate in parallel, processing their respective inputs until a selection process determines which information will enter the global workspace for conscious access.

## Thoughtseeds as Information Processors

Thoughtseeds, in turn, act as the information processors within this GWT framework. They dynamically integrate information from multiple KDs, forming transient coalitions of neural activity that compete for access to the Global Workspace. Each thoughtseed, with its associated core and subordinate attractors, represents a candidate for conscious access, vying to shape the contents of awareness. These attractors can be seen as metastable states, aligning with the brain's pullback attractor dynamics (Friston & Ao, 2012).

## Competition for Dominance

Thoughtseeds compete in a winner-take-all dynamic, where the one that minimizes cumulative expected free energy (EFE)—balancing both epistemic (uncertainty reduction) and pragmatic (goal achievement) considerations (Parr & Friston, 2019)—gains access to the Global Workspace. This process is akin to the "ignition" of a neural assembly (Dehaene & Changeux, 2011), in which a critical threshold of neural activity is reached, leading to a self-sustaining pattern of activation.

Once a thoughtseed becomes dominant, it triggers a shift in global brain activity, reflecting a change in the content of consciousness. This mechanism aligns with the theory of discrete conscious states, where consciousness is seen as a sequence of transient attractor states that emerge through coordinated neural activity (Deco et al., 2013; Dehaene et al., 2003). The winning thoughtseed, after gaining access to the Global Workspace, enables the "broadcast" of its content across the cognitive system, influencing processes such as attention, decision-making, and motor planning, facilitating adaptive behavior (Dehaene, 2014; Deco et al., 2015).

In this way, the global workspace functions as a "broadcasting platform", while KDs act as the "sources" of specialized knowledge, and thoughtseeds serve as the "processors" that dynamically integrate this knowledge to form the content of consciousness (Mashour et al., 2020). Through this mechanism, thoughtseeds shape both perception and action, allowing the cognitive system to respond flexibly to changing environmental demands.

### The Unitary Nature of Conscious Experience

The unitary nature of conscious experience, the subjective feeling of a single, coherent stream of consciousness, emerges from the dominance of a single thoughtseed at any given moment. This dominance reflects the winner-take-all dynamic inherent in the Global Workspace Theory (GWT), ensuring that only one thoughtseed broadcasts its content globally, while other competing thoughtseeds remain inactive or latent. This continuous competition maintains the coherence and integration of conscious experience, preventing fragmented or conflicting streams of awareness from simultaneously entering the workspace (Dehaene & Naccache, 2001).

Thus, thoughtseeds do not merely compete for access to the global workspace but actively contribute to the unitary and seamless nature of conscious experience. Once a thoughtseed gains dominance, it effectively "controls" the global workspace, shaping the current content of consciousness and guiding future actions and decisions. This guidance is achieved through policy selection that balances epistemic (learning) and pragmatic (goal-driven) affordances (Friston et al., 2015).

This cyclical process—of competition, selection, and dominance—underpins the seamless flow of conscious experience, aligning with the brain's overarching goal of minimizing surprise and maximizing adaptive efficiency through active inference.

## Meta-Cognition and Agency in Thoughtseed Dynamics

**Meta-cognition** refers to the ability to monitor, evaluate, and regulate one's own cognitive processes. In the context of thoughtseeds and the Global Workspace Theory (GWT), meta-cognition plays a crucial role in the orchestration of thoughtseed dynamics, influencing which thoughtseed is selected as dominant and how it interacts with broader agent-level goals and policies (Desimone & Duncan, 1995). This aligns with Dehaene's conscious access hypothesis, which posits that conscious access to information is a selective process mediated by prefrontal cortex activity (Dehaene & Naccache, 2001).

## Meta-Cognitive Regulation of Thoughtseeds

Meta-cognitive processes, potentially implemented by higher-order thoughtseeds, help modulate the activation and competition among lower-level thoughtseeds by dynamically adjusting attention, prediction, and error signals. These higher-order thoughtseeds can be seen as "observers" or "regulators" within the cognitive system, evaluating the relevance and coherence of lower-level thoughtseeds and influencing their access to the Global Workspace.

Thoughtseeds compete for dominance based on their ability to minimize free energy, but meta-cognition introduces a higher level of control by shaping *attention* and *prior expectations*. For instance, attentional control can amplify the precision of certain sensory signals, enhancing the chances of relevant thoughtseeds being selected for conscious access (Feldman & Friston, 2010). The *attentional precision parameter* plays a crucial role in this modulation, affecting how sensitive lower-level thoughtseeds are to sensory evidence during their free energy minimization process. By adjusting this parameter, higher-order thoughtseeds can influence which lower-order thoughtseeds gain prominence in the competition for conscious access. This selection process, in line with Dehaene's work, contributes to the discrete nature of conscious experience, where transitions between dominant thoughtseeds correspond to distinct shifts in conscious content.

Moreover, meta-cognition adjusts the activation threshold that determines which thoughtseeds enter the active thoughtseed pool, influencing the pool’s composition. This threshold can vary based on the current cognitive or emotional state. For example, in states of high arousal or focused attention, the threshold may be lowered, allowing a broader range of thoughtseeds to manifest. Conversely, in states of reduced attentional control, only highly salient thoughtseeds surpass the threshold (Corbetta & Shulman, 2002). The *meta-awareness parameter* enriches this process by representing the higher-order thoughtseed's awareness of its influence on lower-level thoughtseeds and attentional focus. This meta-awareness can be modeled as a probabilistic belief about the degree of opacity in accessing and understanding lower-level thoughtseeds and their associated Knowledge Domains (KDs). It serves as a metacognitive monitoring and control mechanism (Metzinger, 2003; Sandved-Smith et al., 2021).

## Agency and Goal-Directed Thoughtseed Dynamics

**Agency** in this context refers to the ability of the cognitive system to act intentionally and adaptively by selecting the most appropriate thoughtseeds based on its goals. Thoughtseeds, acting as sub-agents within the larger cognitive architecture, contribute to the agent's overarching goals by driving behavior that aligns with the minimization of long-term surprise or free energy (Seth, 2015).

Higher-level thoughtseeds, representing more abstract or long-term goals, act as the *agents of control* that guide the system's overall direction. These thoughtseeds exert a top-down influence on the selection and competition process, ensuring that the selected thoughtseed aligns with global goals and the broader objectives of the agent. For example, a higher-order goal (e.g., finding food) might bias the selection of thoughtseeds that prioritize related affordances within the environment, such as detecting food sources or planning routes to them (Friston, 2013).

## Policies, Goals, and Affordances at the Agent Level

At the agent level, thoughtseeds help shape the perception of policies (strategies), goals, and affordances (opportunities for action) within the environment. These elements guide the living system's adaptive behavior by integrating sensory data with its internal models of the world. The active inference framework posits that the agent constantly selects policies that minimize EFE, balancing *epistemic affordances* (opportunities to reduce uncertainty or gain knowledge) with *pragmatic affordances* (opportunities to fulfill goals or needs) (Friston et al., 2015).

In this view, meta-cognition plays a key role in determining which goals are prioritized and how policies are enacted. As the agent navigates its environment, thoughtseeds help to represent and update predictions about the most relevant actions and affordances. This aligns with the notion of a *hierarchical policy structure*, where lower-level thoughtseeds handle immediate, task-specific actions, and higher-level thoughtseeds manage broader, long-term goals and strategies (Pezzulo et al., 2013). The brain's *ergodic principles* ensure that these higher-level goals are regularly revisited, influencing future behavior and keeping the system aligned with long-term adaptive objectives (Friston et al., 2021).

An illustrative example of a "dog" thoughtseed is discussed in Supplementary Section 8.5.

# 4 Illustrative Mathematical Framework

## Neuronal Packets and Superordinate Ensembles

**State Representation** $\mathbf{x}_\nu(\mathbf{t})$ of a NP $\nu$ is represented by its core attractor $\psi_{\nu-c}$ and subordinate attractors $\psi_{\nu-s_i}$ and their corresponding activation levels $\alpha_{\nu-c}$ and $\alpha_{\nu-s_i}$, ∈ [0, 1]. $\boldsymbol{\omega}_\nu$ represents the object/knowledge encapsulated within $\nu$. $s_\nu$ its state ∈ {0: unmanifested, 1: manifested, 2: activated}.

$$\mathbf{x}_\nu(t) = \{(\psi_{\nu-c}, \alpha_{\nu-c}(t)), (\psi_{\nu-s_1}, \alpha_{\nu-s_1}(t)), \ldots, (\psi_{\nu-s_n}, \alpha_{\nu-s_n}(t)), \boldsymbol{\omega}_\nu, s_\nu\} \quad (1)$$

**Generative Model** of a NP $\nu$ describes how its internal states $\mathbf{x}_\nu(\mathbf{t})$ aids in generating sensory predictions $\mathbf{s}_\nu$ and potentially, actions $\mathbf{a}_\nu$. Given the internal model parameters $\theta_\nu$, state of the parent KD $\mathbf{x}_{\mathcal{K}^*}$, encapsulated knowledge $\boldsymbol{\omega}_\nu$ and the NP state $s_\nu$. A NP can be further classified as a $type_{NP}$(sensory, active, or internal) for additional granularity.

$$p(\mathbf{x}_\nu, \mathbf{a}_\nu, \mathbf{s}_\nu | \theta_\nu, \mathbf{x}_{\mathcal{K}^*}, \boldsymbol{\omega}_\nu, s_\nu) \tag{2}$$

**Free Energy Minimization** of a NP. Each NP minimizes its free energy $F_\nu$, by balancing two opposing pressures: Complexity and Accuracy.

$$F_\nu = \underbrace{D_{KL}\left[q\left(\mathbf{x}_\nu\right) || p\left(\mathbf{x}_\nu | \mathbf{a}_\nu, \mathbf{s}_\nu, \theta_\nu, \mathbf{x}_{\mathcal{K}^*}, \boldsymbol{\omega}_\nu, s_\nu\right)\right]}_{\text{Complexity}} - \underbrace{\mathbb{E}_{q(\mathbf{x}_\nu)}\left[\ln p\left(\mathbf{a}_\nu, \mathbf{s}_\nu | \mathbf{x}_\nu, \theta_\nu, \mathbf{x}_{\mathcal{K}^*}, \boldsymbol{\omega}_\nu, s_\nu\right)\right]}_{\text{Accuracy}} \tag{3}$$

where:

- **Complexity:** This term, represented by the *Kullback-Leibler (KL) divergence*, $D_{KL}$ pushes the NP to minimize the complexity of its internal state representation, $\mathbf{x}_\nu$. $D_{KL}$ measures the difference between the NP's approximate posterior distribution, $q$ term, and it's true posterior distribution, $p$ term. By minimizing this divergence, the NP seeks the simplest possible internal state that remains consistent with its sensory inputs, actions, model parameters, the state of its parent KD, its encapsulated knowledge, and its current activation state.
- **Accuracy:** This term, represented by the *negative log-likelihood*, reflects the NP's drive to accurately predict its active and sensory states, $\mathbf{a}_\nu$ and $\mathbf{s}_\nu$. This involves maximizing the expected log-likelihood of its generative model. By maximizing this term, the NP strives to generate actions that lead to expected sensory inputs, minimizing prediction error and surprise. This process is influenced by the state of its parent KD, its own internal state, its encapsulated knowledge, and its current activation state.

**State Representation** $\mathbf{x}_\varepsilon(t)$ of a **Superordinate Ensemble (SE)** $\boldsymbol{\varepsilon}$ is a functional unit within a KD that integrates information from multiple NPs. It represents a higher-level pattern of neural activity that emerges from the coordinated activity of its constituent NPs.

$$\mathbf{x}_\varepsilon(t) = (\boldsymbol{\psi}_\varepsilon, \boldsymbol{\alpha}_\varepsilon(t), \boldsymbol{\omega}_\varepsilon, s_\varepsilon, v(t), r(t)) \tag{4}$$

$\psi_{\varepsilon}$ and $\boldsymbol{\alpha}_{\varepsilon}(t)$ represents the set of attractors (core and subordinates) and the corresponding activation levels of SE $\varepsilon$. $\boldsymbol{\omega}_{\varepsilon}$ represents the encapsulated knowledge structure, valence $v(t)$ ∈ [-1, 1] (unpleasantness/pleasantness) and arousal $r(t)$ ∈ [0, 1] of SE $\varepsilon$.

## Knowledge Domains (KDs)

**State Representation** $\mathbf{x}_{\mathcal{K}}(t)$ of a **KD** $\mathcal{K}$ is a collection of SEs that together represent a broader domain of knowledge or expertise. It acts as a repository for integrated knowledge shaped by the interactions and relationships between its constituent SEs $\varepsilon_i$.

$$\mathbf{x}_{\mathcal{K}}(t) = \{\mathbf{x}_{\varepsilon_1}(t), \mathbf{x}_{\varepsilon_2}(t), ..., \mathbf{x}_{\varepsilon_n}(t)\} \qquad (5)$$

**Generative Model** of a **KD** $\mathcal{K}$ describes how the KD predicts its active states $\mathbf{a}_{\mathcal{K}}$, sensory states $\mathbf{s}_{\mathcal{K}}$, internal state $\mathbf{x}_{\mathcal{K}}$ (which includes the states of its constituent SEs), content of consciousness $\kappa_{\mathcal{K}}$ which is hypothesized to capture the *qualia* (Chalmers, 1995) or subjective experience associated with the KD's activation, and affective states (valence $v_{\mathcal{K}}$ and arousal $r_{\mathcal{K}}$), given its model parameters $\theta_{\mathcal{K}}$ and the state of its parent thoughtseed $\mathbf{x}_{\mathcal{T}^*}$. It formalizes the KD's internal model of the world and its role in shaping subjective experience and guiding behavior within the thoughtseed framework.

$$p(\mathbf{a}_{\mathcal{K}}, \mathbf{s}_{\mathcal{K}}, \mathbf{x}_{\mathcal{K}}, \kappa_{\mathcal{K}}, v_{\mathcal{K}}, r_{\mathcal{K}} | \theta_{\mathcal{K}}, \mathbf{x}_{\mathcal{T}^*}) \qquad (6)$$

**Free Energy Minimization** $F_{\mathcal{K}}$ of a **KD** $\mathcal{K}$ effectively balances the need for a parsimonious internal representation with the need for accurate predictions about its interactions with the world. This process ensures that the KD contributes to adaptive behavior while maintaining an efficient and coherent representation of its knowledge domain.

$$F_{\mathcal{K}} = \underbrace{D_{KL}\left[q\left(\mathbf{x}_{\mathcal{K}}\right) || p\left(\mathbf{x}_{\mathcal{K}} | \mathbf{a}_{\mathcal{K}}, \mathbf{s}_{\mathcal{K}}, \kappa_{\mathcal{K}}, v_{\mathcal{K}}, r_{\mathcal{K}}, \theta_{\mathcal{K}}, \mathbf{x}_{\mathcal{T}^*}\right)\right]}_{\text{Complexity}}$$

$$-\underbrace{\mathbb{E}_{q(\mathbf{x}_{\mathcal{K}})}\left[\ln p\left(\mathbf{a}_{\mathcal{K}}, \mathbf{s}_{\mathcal{K}}, \kappa_{\mathcal{K}}, v_{\mathcal{K}}, r_{\mathcal{K}} | \mathbf{x}_{\mathcal{K}}, \theta_{\mathcal{K}}, \mathbf{x}_{\mathcal{T}^*}\right)\right]}_{\text{Accuracy}} \quad (7)$$

where:

- **Complexity:** The Complexity term considers how well the KD's internal state representation $\mathbf{x}_{\mathcal{K}}$ explains its actions $\mathbf{a}_{\mathcal{K}}$, sensory inputs $\mathbf{s}_{\mathcal{K}}$, content of consciousness $\kappa_{\mathcal{K}}$, valence $v_{\mathcal{K}}$, arousal $r_{\mathcal{K}}$, model parameters $\theta_{\mathcal{K}}$, and the state of its parent thoughtseed $\mathbf{x}_{\mathcal{T}^*}$.
- **Accuracy:** The Accuracy term considers how well the KD's generative model predicts its actions, sensory inputs, content of consciousness, valence, and arousal, given its internal state and the influence of its parent thoughtseed.
- Weighted difference between the predicted and actual valence and arousal associated with a KD can also be added as shown below:
  $\lambda_v(\mathbf{v}_{\mathcal{K}} - \hat{\mathbf{v}}_{\mathcal{K}})^2 + \lambda_r(\mathbf{r}_{\mathcal{K}} - \hat{\mathbf{r}}_{\mathcal{K}})^2$

## Thoughtseeds

Thoughtseeds can be conceptualized as information processing units competing for access to the global workspace, the proposed neural correlate of consciousness.

**State Representation** $\mathbf{x}_{\mathcal{T}}(t)$ of a **Thoughtseed** $\mathcal{T}$ is represented by its core attractor $\psi_{\mathcal{T}-c}$ and subordinate attractors $\psi_{\mathcal{T}-s_i}$ and their corresponding activation levels $\alpha_{\mathcal{T}-s_i}$ and $\alpha_{\mathcal{T}-s_i}$, ∈ [0, 1]. $\mathcal{K}_{\mathcal{T}}$ represents the encapsulated knowledge structures within $\mathcal{T}$. $s_{\mathcal{T}}$ its state ∈ {0: unmanifested, 1: manifested-inactive, 2: manifested -active, 3:manifested-dominant, 4:dissipated}. $v_{\mathcal{T}}$ ∈ [-1, 1] (unpleasantness /pleasantness) and arousal $r_{\mathcal{T}}$ ∈ [0, 1] of $\mathcal{T}$.

$$\mathbf{x}_{\mathcal{T}}(t) = \{(\psi_{\mathcal{T}-c}, \alpha_{\mathcal{T}-c}(t)), (\psi_{\mathcal{T}-s_1}, \alpha_{\mathcal{T}-s_1}(t)), \ldots, (\psi_{\mathcal{T}-s_m}, \alpha_{\mathcal{T}-s_m}(t)), \mathcal{K}_{\mathcal{T}}, s_{\mathcal{T}}(t), v_{\mathcal{T}}(t), r_{\mathcal{T}}(t)\} \quad (8)$$

**Activation Level of a Thoughtseed** $\mathcal{T}$ is represented by a weighted combination of the brain's current state in the thoughtseed's core attractor $\psi_{\mathcal{T}-c}$ or subordinate attractors $\psi_{\mathcal{T}-s_i}$.

$$\alpha_{\mathcal{T}}(t) = w_c \cdot p(\psi_{\mathcal{T}-c} \mid \mathbf{x}_{\mathcal{T}}(t)) + \sum_{j=1}^{m} w_{s_j} \cdot p(\psi_{\mathcal{T}-s_j} \mid \mathbf{x}_{\mathcal{T}}(t)) \tag{9}$$

The activation level represents a thoughtseed's "bid" for dominance. The thoughtseed with the highest activation level is most likely to enter the Global Workspace, shaping conscious experience.

**Generative Model of a Thoughtseed:** predicts its own future state $\mathbf{x}_{\mathcal{T}}(t+1)$, encompassing changes in its core attractor, associated knowledge, and emotional tone while simultaneously generating the content of consciousness $\mathbf{c}_{\mathcal{T}}(t)$.

$$p(\mathbf{x}_{\mathcal{T}}(t+1), \mathbf{c}_{\mathcal{T}}(t) | \mathbf{x}_{\mathcal{T}}(t), \mathbf{x}_{\mathcal{K}}(t), v_{\mathcal{T}}(t), a_{\mathcal{T}}(t), \theta_{\mathcal{T}}, \mathbf{u}, \sigma, \mathbf{A}, \mathbf{\Pi}) \tag{10}$$

This prediction is conditioned upon its current state $\mathbf{x}_{\mathcal{T}}(t)$, the state of the knowledge domains $\mathbf{x}_{\mathcal{K}}(t)$ it draws from, and the valence $v_{\mathcal{T}}$ and arousal $r_{\mathcal{T}}$ associated with those domains. It also considers sensory input $\mathbf{u}$ and its salience $\sigma$, perceived affordances for action $\mathbf{A}$, internal model parameters $\theta_{\mathcal{T}}$, and policies guiding action selection $\mathbf{\Pi}$. This dual prediction allows the thoughtseed to anticipate its own evolution and influence on cognition and behavior while actively shaping conscious experience within the broader cognitive landscape.

**Free Energy Minimization of a Thoughtseed** is a measure comprising both complexity and accuracy. The complexity term reflects a thoughtseed's propensity to reduce uncertainty and surprise, compelling it to actively seek information within the global workspace. This can be interpreted as an inclination towards acquiring information from the ongoing broadcast, by predicting its next state $\mathbf{x}_{\mathcal{T}}(t+1)$ and generating the content of consciousness $\mathbf{c}_{\mathcal{T}}(t)$ . Conversely, the accuracy term compels a thoughtseed to maximize the evidence for its internal model, prompting it to influence the content of the global broadcast. This can be understood as a tendency towards contributing information to the workspace.

$$F_{\mathcal{T}} = \underbrace{D_{KL}[q(\mathbf{x}_{\mathcal{T}}(t+1), \mathbf{c}_{\mathcal{T}}(t))||p(\mathbf{x}_{\mathcal{T}}(t+1), \mathbf{c}_{\mathcal{T}}(t)|\mathbf{x}_{\mathcal{T}}(t), \mathbf{x}_{\mathcal{K}}(t), v_{\mathcal{T}}(t), a_{\mathcal{T}}(t), \theta_{\mathcal{T}}, \mathbf{u}, \sigma, \mathbf{A}, \mathbf{\Pi})]}_{\text{Complexity}}$$

$$- \underbrace{\mathbb{E}_{q(\mathbf{x}_{\mathcal{T}}(t+1), \mathbf{c}_{\mathcal{T}}(t))}[\ln p(\mathbf{s}(t), \mathbf{a}(t)|\mathbf{x}_{\mathcal{T}}(t), \mathbf{x}_{\mathcal{K}}(t), \mathbf{c}_{\mathcal{T}}(t), \theta_{\mathcal{T}})]}_{\text{Accuracy}} \quad (11)$$

## Active Thoughtseeds Pool

The Thoughtseeds Network comprises numerous manifested thoughtseeds vying for conscious expression, but only a select few actively contribute to the stream of consciousness.

$$\mathcal{P}_{\textbf{active}}(t-\Delta t, t) = \mathbf{x}_{\mathcal{T}i}(t')|\alpha_{\mathcal{T}i-c}(t') > \Theta_{activation}(t'), t' \in (t-\Delta t, t] \quad (12)$$

This selection is mediated by a global activation parameter $\Theta_{activation}(t)$, influenced by the state of consciousness and arousal levels, serving as a dynamic threshold for entry into the "active thoughtseed pool." This parameter regulates the "gain" of the Global Workspace, determining the selectivity and sensitivity of conscious awareness. This pool $\mathcal{P}_{\textbf{active}}$, fluctuating within brief intervals, represents the ensemble of thoughtseeds competing for dominance within the Global Workspace.

## Meta-cognition

### Dominant Thoughtseed and Content of Consciousness

**Selection of the Dominant Thoughtseed:** The competition among thoughtseeds in the active pool $\mathcal{P}_{\textbf{active}}$ follows a winner-take-all dynamic. The thoughtseed that most effectively minimizes its free energy emerges as the dominant thoughtseed $\mathcal{T}^*$. It gains privileged access to the Global Workspace, broadcasting its information widely and influencing a broad range of cognitive processes.

$$\mathcal{T}^*(t-\Delta t, t) = \underset{\mathcal{T}_i \in \mathcal{P}\mathbf{active}(t-\Delta t, t)}{\operatorname{argmin}} \sum_{t'=t-\Delta t}^{t} F_{\mathcal{T}_i}(t') \tag{13}$$

This privileged access allows the dominant thoughtseed to shape conscious experience, guide attention, and direct action, effectively orchestrating thought and behavior. This selection process can also be chosen via a cumulative Effective Free Energy (EFE) minimization. We additionally hypothesize that for long-term or deliberative planning Generalized Free Energy (GFE) minimization could be a better candidate for dominant thoughtseed selection.

### Unitary Nature of the Content of Consciousness

The content $\mathcal{C}(t)$ in the Global Workspace at any moment is shaped a single, dominant thoughtseed reflecting the unitary nature of conscious experience. The identification of the dominant thoughtseed index $i$ is articulated in Eq 13.

$$\mathcal{C}(t) = \mathbf{c}_i^*(t-\Delta t, t) \tag{14}$$

### Influence of Higher-order Thoughtseeds on Lower-order Thoughtseeds

Higher-order thoughtseeds with meta-awareness $m$, exert influence on lower-level thoughtseeds by modulating their attentional precision $\gamma$. This influence is captured by Eq 15, with $\{\mathbf{x}_\tau\}_{\tau \in \mathcal{T}}$ denotes the internal states of competing lower-level thoughtseeds. $\mathbf{\Pi}$encapsulates the set of policies, $\mathbf{G}$ represents the goals pursued by the higher-order thoughtseed, conditioned on model parameters $\theta_m$, prior beliefs $\mathbf{u}$ and the active thoughtseeds in $\mathcal{P}_{\mathbf{active}}$.

$$p(\gamma, \{\mathbf{x}_\tau\}_{\tau \in \mathcal{T}}, m, \mathbf{\Pi}, \mathbf{G} | \theta_m, \mathbf{u}, \{i \in \mathcal{P}_{\mathbf{active}}\}) \tag{15}$$

## Goals, Policies and Affordances at Agent Level

**Global Goals:** The agent's goals $\mathbf{G}_{agent}$ can be described as a function of the characteristic states $\chi_i^*$ within the brain's pullback attractor landscape. These states are revisited frequently (above a certain threshold $\theta_{freq}$) and demonstrate a level of stability or persistence $S_i$. This goal function is represented as:

$$g_{agent} : \{(\chi_i^*, S_i) | f_i > \theta_{freq}\} \to \mathbf{G}_{agent}(t) \quad (16.1)$$

**Global Policies:** Once the agent's goals are established, the selection of appropriate policies (or action plans) to achieve these goals is guided by the knowledge and beliefs contained within various Knowledge Domains (KDs), along with the context and perceived affordances of the environment. The agent-level policies $\mathbf{\Pi}_{agent}$ are defined as a function of the states of all KDs $\{\mathbf{x}_K(t)\}_{K \in \mathcal{K}}$ and the agent's global goals:

$$h_{agent} : \{\mathbf{x}_K(t)\}_{K \in \mathcal{K}} \times \mathbf{G}_{agent}(t) \to \mathbf{\Pi}_{agent}(t) \quad (16.2)$$

The functions $g_{agent}(t)$ and $h_{agent}$ encapsulate the complex processes involved in deriving goals and policies from the interplay of characteristic states, knowledge domains, and the Agent's internal dynamics.

**Global Affordances:** The global affordances at an Agent level are described via mapping functions $k_{e-agent}$ and $k_{p-agent}$ respectively.

$$k_{e-agent} : \{\mathbf{x}_K(t)\}_{K \in \mathcal{K}} \times \mathbf{s} \to \mathbf{A}_{epistemic-agent}(t) \quad (16.3)$$

$$k_{p-agent} : \{\mathbf{x}_K(t)\}_{K \in \mathcal{K}} \times \mathbf{s} \times \mathbf{G}_{agent}(t) \to \mathbf{A}_{pragmatic-agent}(t) \quad (16.4)$$

While thoughtseed-specific affordances are more context-dependent and tailored to the individual thoughtseed's knowledge and goals, global affordances provide a broader view of the possible actions available to the agent, considering its overarching objectives and the integrated knowledge across all KDs.

### Agent-Level Free Energies: GFE, VFE and EFE

**Agent-level GFE** is calculated as the sum of the current Variational Free Energy (VFE) and the expected value of the Expected Free Energy (EFE) under the agent's current policy. This captures both the current surprise or uncertainty about the agent's sensory observations $\mathbf{s}$ and the anticipated uncertainty related to future potential actions $\mathbf{a}$.

$$GFE_{agent}(t) = VFE_{agent}(t) + \mathbb{E}_{q(\mathbf{s},\mathbf{a}|\Pi_{agent}(t))}[EFE_{agent}(t+1)] \quad (17.1)$$

**Agent-level VFE** is calculated as sum of VFEs of active thoughtseeds in $\mathcal{P}_{\mathbf{active}}(t)$ plus other factors contributing to the agent's overall surprise at that instant.

$$VFE_{agent}(t) = \sum_{i \in \mathcal{P}_{\mathbf{active}}(t)} VFE_i(t) + \dots \quad (17.2)$$

**Agent-level EFE** is calculated as the sum of EFEs of active thoughtseeds in $\mathcal{P}_{\mathbf{active}}(t)$ weighted by their respective activation levels $\alpha_i(t)$. This allows the Agent to evaluate potential outcomes of different policies, considering both their epistemic(informational gain) and pragmatic (goal-oriented) values, and select the policy $\Pi_{agent}(t)$ that minimizes overall surprise.

$$EFE_{agent}(\mathbf{\Pi}_{agent}, t) = \sum_{i \in \mathcal{P}_{\mathbf{active}}(t)} \alpha_i(t) \cdot EFE_i(\mathbf{\Pi}_i, t) \quad (17.3)$$

# 5. Discussion and Future Directions

## Towards a General Theory of Embodied Cognition

The Thoughtseed Framework, building upon the concept of neuronal packets (Yufik, 2013, 2019; Yufik & Friston, 2016; Ramstead et al., 2021) and the and the Free Energy Principle (FEP) (Friston, 2010), offers a promising foundation for a general theory of embodied cognition (Allen & Friston, 2018; Foglia & Wilson, 2013; Pezzulo et al., 2024). Within this framework, thoughtseeds—emergent, higher-order constructs—arise from the coordinated activity across knowledge domains (KDs). These thoughtseeds represent coherent patterns of neural activity that serve as agentic constructs, driving action, perception, and cognitive function. This aligns closely with the Global Workspace Theory (GWT) (Baars, 1997; Dehaene & Changeux, 2011), where thoughtseeds can be seen as high-salience representations that gain access to the global workspace, thereby influencing conscious experience. Furthermore, the thoughtseed framework aligns with the notion of discrete thought processes, where each dominant thoughtseed represents a distinct cognitive moment or 'atom' of conscious experience.

The thoughtseed framework introduces the novel concept of thoughtseeds as self-organizing units with their own internal states and dynamics, which sets it apart from other cognitive theories. These self-organizing dynamics, driven by the principles of active inference, allow thoughtseeds to actively engage with the environment, predict future states, and shape the ongoing flow of conscious experience. This active and embodied nature of thoughtseeds provides a unique perspective on how cognition emerges from the interaction between the living system and its environment.

Cognition, as seen in this framework, is not merely an internal process. It is inherently *embodied* and situated, deeply rooted in the interactions between the living system's internal states, its body, and the environment (Thompson, 2007). Thoughtseeds emerge from this reciprocal relationship and shape the living system's perception and decision-making processes by forming predictions, selecting actions, and integrating sensory feedback.

## The Hierarchical Structure of the Thoughtseed Framework

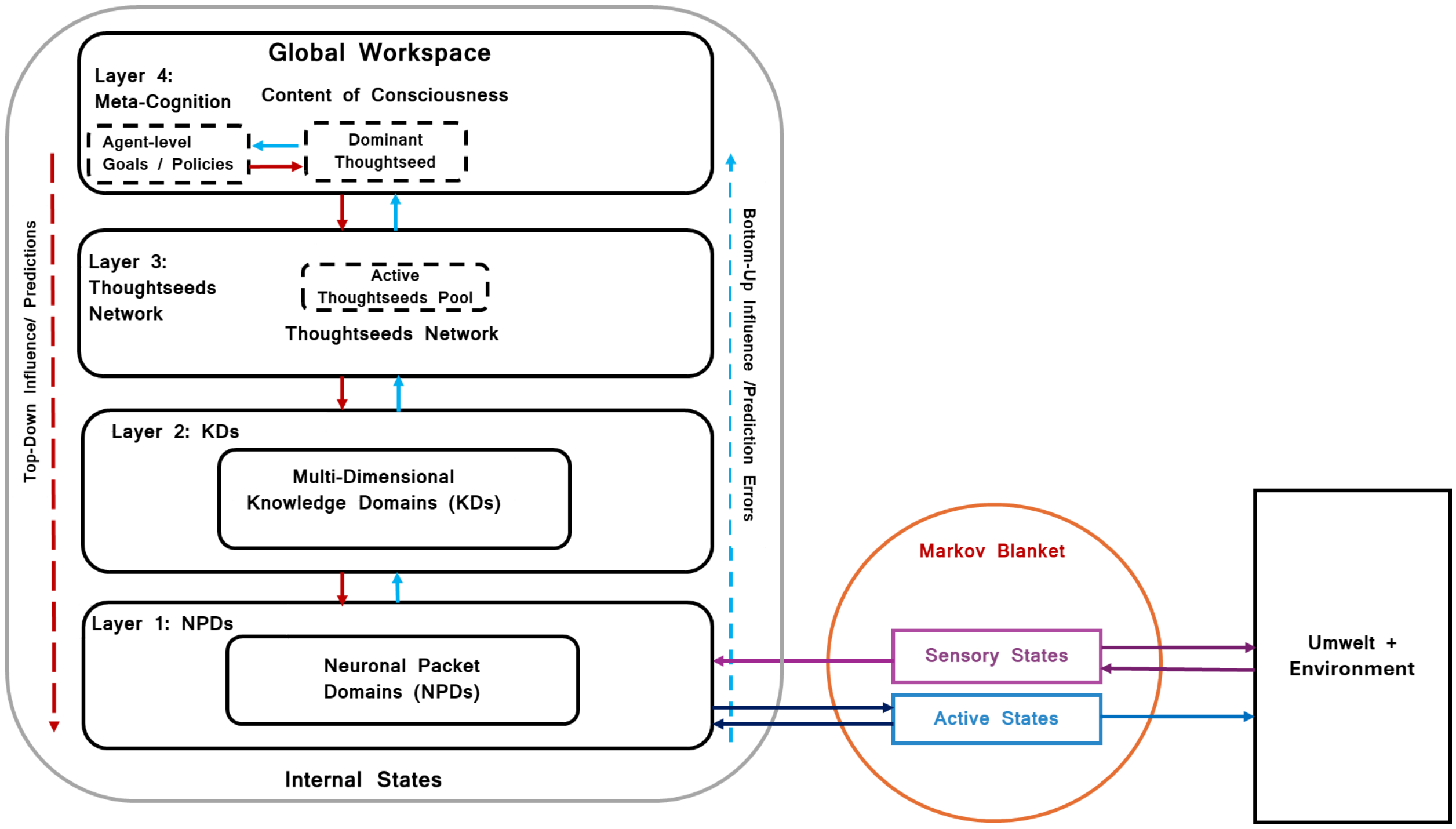

**Figure 3. Illustrative Thoughtseed Framework.** This diagram provides a high-level conceptual overview of the internal states within the Thoughtseed Framework, structured across four nested levels. It employs an active inference formalism framed within the principles of Global Workspace Theory.

The Thoughtseed Framework proposes a hierarchical organization of cognitive processes (see Figure 3), offering a mechanistic explanation for how the brain organizes and manages complex functions such as *decision-making, problem-solving*, and *planning*. These processes can be understood within the Global Workspace Theory (GWT) paradigm, where information from various subsystems becomes globally accessible to the living system when it reaches the level of conscious awareness. Below is an outline of the hierarchy:

1. **Neuronal Packet Domains (NPDs):** NPDs form the foundation of the framework, consisting of groups of neurons that process sensory input or generate potential actions. Within the GWT, these subsystems provide the foundational, localized processing of information. NPDs can be divided into domains responsible for sensory processing, motor planning, and internal state regulation, contributing raw data for higher-order representations.

2. **Knowledge Domains (KDs):** KDs represent large-scale networks that integrate knowledge, memories, and conceptual categories. They serve as structured repositories from which thoughtseeds draw. Within the GWT, KDs act as functional modules that hold specialized knowledge, accessible to the global workspace when relevant. For example, recognizing a dog would involve KDs related to visual processing, auditory associations (barking), and emotional contexts (companionship).
3. **Thoughtseeds Network:** Thoughtseeds represent dynamic, self-organizing entities that emerge from the coordinated activity within and across KDs. In the GWT context, thoughtseeds represent potential candidates for entry into the global workspace, where they can shape conscious experience and direct behavior. Each thoughtseed is a coherent pattern of neural activity associated with specific concepts, percepts, or possible actions. Thoughtseeds act as cognitive sub-agents, actively generating predictions and selecting actions to minimize free energy, constantly updating based on sensory feedback.
4. **Meta-Cognition, Dominant Thoughtseed and Higher-Order Thoughtseeds:** This level involves higher-order thoughtseeds that modulate attention and guide goal-directed behavior. Higher-order thoughtseeds reflect the agent's global goals, influencing which lower-level thoughtseeds gain prominence within the global workspace. In GWT terms, these higher-order constructs govern the selection of conscious content by modulating attentional precision and prioritizing relevant information. This helps shape global policies that determine the living system's actions in a goal-oriented manner. The selection of a dominant thoughtseed can be seen as a discrete event, leading to a distinct shift in conscious content.

The culmination of these hierarchical processes is the projection of content onto the Global Workspace, the metaphorical stage of conscious experience. The dominant thoughtseed at any moment, selected through competitive processes based on free energy minimization and influenced by meta-cognitive processes, shapes the conscious content. This content includes sensory input, memories, concepts, and potential actions, reflecting the living system's current understanding and interaction with the environment. The dynamic interplay between thoughtseeds and knowledge domains creates the fluid "stream of thoughts" that manifests as the *content of consciousness*, guiding adaptive behavior and decision-making.

## Limitations and Future Research Directions

Investigating the neural mechanisms behind thoughtseed formation necessitates a multi-scale approach, spanning from the co-activation of neurons to the emergence of stable Markov blankets embodying distinct attractor states. A major challenge lies in developing biologically plausible computational models of spiking neural networks that incorporate synaptic plasticity and homeostatic mechanisms. Such models could simulate the self-organization of thoughtseeds, allowing researchers to track the evolution of attractor state stability, precision, and how these states are bound within nested Markov blankets to foster thoughtseed resilience and predictive (Yufik, 2019; Litwin-Kumar & Doiron, 2012; Hipólito et al., 2021).

These computational models could also help identify conditions under which thoughtseeds transition between states, reflecting attractor dynamics seen in models of memory (Boscaglia et al., 2023; Parr, et al., 2021). Potential neural signatures, such as spatiotemporal patterns of activity, connectivity motifs, or oscillatory dynamics, could indicate the presence of thoughtseeds (Buzsáki, 2006; Fries, 2015; Sporns, 2011). . Early research could focus on specific cognitive domains like object permanence or numerical cognition to observe how thoughtseeds manifest in these processes.

### Key Limitations

1. **Metastability of Thoughtseeds**: Thoughtseed dynamics are inherently metastable, with rapid transitions between states. This makes it difficult to isolate core attractors representing stable knowledge or behaviors, posing challenges in capturing and measuring the dynamic properties of thoughtseeds empirically (Rabinovich et al., 2008; Tognoli & Kelso, 2014; Deco et al., 2016, 2017).

2. **Mapping Distributed Neural Activity**: Given the complexity of the brain's distributed networks, linking specific patterns of neural activity to cognitive processes (e.g., thoughtseeds) is fraught with difficulty. Misinterpretation could arise without a clear framework that maps distributed signals to well-defined cognitive domains (deCharms & Zador, 2000; Rué-Queralt, et al., 2021).
3. **Hierarchical Complexity**: The nested Markov blanket structure of thoughtseeds introduces additional complexity in understanding how higher- and lower-order processes interact. This complicates experimental design and analysis, especially when attempting to link abstract thought processes to neurobiological signatures (Friston, 2020; Kirchhoff et al., 2018).

## Future Directions

Future research should focus on several key areas to further develop and validate the thoughtseed framework:

- **Computational Modeling:** Future research should prioritize the development and empirical validation of computational models that simulate thoughtseed dynamics within the Global Workspace. These models could utilize hierarchical nested Markov-blanketed structures and leverage frameworks like the intrinsic ignition model (Deco & Kringelbach, 2017), but adapted to use thoughtseeds as the competing units. By incorporating thoughtseed internal structure, active inference mechanisms, and EFE minimization for the "ignition" of the Global Workspace, these models can explore how parameters like activation thresholds and attentional precision influence the emergence of discrete conscious states and dominant thoughtseed selection. Furthermore, these models could plausibly generate testable predictions about the neural correlates of thoughtseed transitions, allowing for comparisons with empirical data from neuroimaging techniques like EEG or fMRI and findings on the ignition of neural assemblies (Dehaene, 2014).
- **Cognitive Development:** Investigate how the thoughtseed framework can be applied to understand cognitive development. This could involve exploring how thoughtseed dynamics change across the lifespan and contribute to learning, memory, and the acquisition of complex cognitive skills. Early research could focus on specific cognitive domains like object permanence (Piaget, 1954) or numerical cognition (Nieder, 2017) to observe how thoughtseeds manifest in these processes. The development of thoughtseeds may also be investigated in the context of the emergence of discrete symbols and language (Dehaene et al., 2022; Dehaene, 2020).

- **Clinical Applications:** Explore the potential applications of the thoughtseed framework in understanding and treating disorders of attention. This could involve investigating how disruptions in thoughtseed dynamics contribute to specific clinical conditions and whether interventions targeting thoughtseed regulation can improve cognitive function.

By pursuing these research directions, the thoughtseed framework can be further refined and validated, leading to a deeper understanding of cognition, consciousness, and their underlying neural mechanisms.

# 6. Conclusion

The thoughtseed framework offers a novel and biologically plausible model for understanding the emergence and organization of embodied cognition. By integrating evolutionary principles, active inference, and the Global Workspace Theory, the framework provides a comprehensive account of how cognitive processes arise from the dynamic interplay of neuronal packets, knowledge domains, and higher-order thoughtseeds. The framework's emphasis on self-organization, competition, and free energy minimization sheds light on the mechanisms underlying thought formation, conscious experience, and adaptive behavior.

The hierarchical structure of the framework, with its nested Markov blankets and reciprocal message passing, provides a compelling explanation for the brain's ability to integrate information across different levels of representation and generate coherent cognitive states. The concept of thoughtseeds as self-organizing units of embodied knowledge offers a promising new perspective on the nature of thought and its role in shaping conscious experience.

While still in its theoretical stages, the thoughtseed framework provides a fertile ground for future research, with potential applications in understanding cognitive development, psychopathology, and artificial intelligence. By continuing to explore the dynamics of thoughtseeds and their interactions within the Global Workspace, we can gain a deeper understanding of the fundamental principles that govern cognition and consciousness.

# 7. Conflict of Interest

The authors declare that the research was conducted in the absence of any commercial or financial relationships that could be construed as a potential conflict of interest.

# 8. Author Contributions

Conceptualization and Mathematical Formalism, Original Drafting: Prakash; Editing: All authors

# 9. Supplementary Section

## Illustrative Example: The Visual Object Recognition KD

The visual object recognition Knowledge Domain (KD) illustrates the complex nature of KDs and their relationship with Neuronal Packet Domains (NPDs). This KD encompasses the hierarchical processing of visual features, progressing from basic sensory inputs to complex object representations. Additionally, it integrates information from various modalities, such as auditory and tactile inputs, as well as semantic and emotional associations stored in memory. This multi-modal integration facilitates a rich and nuanced understanding of objects that transcends mere visual appearance. For instance, recognizing a dog involves not only processing its visual features but also associating it with characteristic sounds (barking), tactile sensations (fur), and emotional connotations (e.g., loyalty, companionship).

### Hierarchical Processing within the Visual Object Recognition KD

Within the visual object recognition KD, various processing systems are hypothesized to operate in a hierarchical manner, contributing to the construction of a comprehensive visual experience. The primary visual cortex (V1) acts as an initial processing stage, extracting basic features such as edges and orientations from retinal input. The ventral visual stream, encompassing areas like V4 and the lateral occipital complex (LOC), functions to integrate these features into representations of shapes and objects (Grill-Spector et al., 2001). Higher-order regions, such as the inferotemporal cortex (IT), further synthesize this information to recognize complex objects like faces (FFA) or tools (EBA) (Kanwisher et al., 1997). The efficiency of this hierarchical processing can be attributed to a sparse representation of visual information, leading to robust encoding of visual features (Olshausen & Field, 2004; Herzog & Clarke, 2014).

### KD as Integrative Framework

The visual object recognition KD serves as an integrative framework, binding outputs from these various processing systems into coherent percepts. It organizes and contextualizes the information derived from different sensory modalities, allowing for meaningful interpretations of the stimuli. The specific content represented within this framework is dynamically shaped by the interaction between activated KDs, the living system's goals and expectations, and the saliency of environmental stimuli.

This binding process, which leads to the emergence of a coherent percept, involves synchronization of neural activity across different processing systems (Singer, 1999; Palacios, 2019), and the formation of superordinate ensembles that encompass multiple neural representations. The enhancement of firing rates in specific neurons may facilitate the binding of information, creating a more integrated perceptual experience (Roelfsema, 2023). The dynamic and context-dependent nature of this process enables flexible and adaptive perception, allowing the living system to respond effectively to the ever-changing demands of its environment.

### Nested Markov Blankets and Hierarchical Representation

Each neural representation within the visual object recognition KD is hypothesized to possess its own Markov blanket, forming a nested hierarchy of representations. This structure allows for the integration and exchange of information across multiple levels of abstraction. The information encoded at each level is progressively coarse-grained, moving from fine-grained sensory details at lower levels to more abstract and categorical representations at higher levels (Hipólito et al., 2021).

## Illustrative Example: "Dog" Thoughtseed

The "dog" thoughtseed serves as an illustrative example of how thoughtseeds emerge and function within thoughtseeds framework, representing an individual's integrated knowledge, beliefs, and associated behaviors related to dogs (Miklósi, 2014). This encapsulation forms a complex and multifaceted concept within the brain's internal model, which can be understood through the lens of Global Workspace Theory (GWT).

### A Generative Model Perspective of "Dog" Thoughtseed

**Scenario:** You are walking in a park, surrounded by various stimuli, including trees, birds chirping, and other people. Suddenly, you spot a Golden Retriever playing fetch.

**Level 1: Neuronal Packet Domains (NPDs) - The Projectors**

- **Sensory NPDs:**
    - **Visual NPD:** The sight of the Golden Retriever triggers the activation of various neural packets (NPs) within the visual NPD, encoding features like its golden fur, wagging tail, and playful movements. This sensory information is transformed into a format suitable for conscious access and interpretation.
    - **Auditory NPD:** The sound of the dog barking activates NPs within the auditory NPD, contributing to the overall perception of the dog.
    - **Other Sensory NPDs:** If you were to pet the dog or smell it, the somatosensory and olfactory NPDs would also become active, enriching the sensory representation of the dog.

- **Active State NPDs:**
  - **Motor NPDs:** The sight of the dog playing fetch triggers potential actions within the motor NPDs, such as reaching out to pet the dog or throwing a ball. These potential actions are represented as motor plans that are prepared for execution.
- **Internal State NPDs:**
  - These NPDs integrate information from the sensory and active NPDs, contributing to the formation of higher-order representations within relevant Knowledge Domains (KDs). They recognize the combination of visual and auditory features as indicative of a "dog," activating the corresponding SEs within the "animal" KD.

**Level 2: Knowledge Domains (KDs) - The Content Generators**

- **"Animal" KD:** This KD contains Superordinate Ensembles (SEs) representing various animals, their characteristics, and behaviors. Activating the "dog" SE within this KD provides a rich conceptual framework about dogs, their typical actions (e.g., barking, playing), and their relationships with humans (e.g., as pets). This information can be accessed by conscious awareness, becoming part of the global workspace.
- **Other Relevant KDs:**
  - **"Pet" KD:** Provides context about the dog's social role and potential interactions, shaping the observer's expectations and actions.
  - **"Emotion" KD:** Contributes to the affective experience associated with the dog, generating feelings of joy, excitement, or even fear, depending on past encounters.
  - **"Dog Encounters" KD:** Activates memories of past encounters with dogs, enriching the representation of the current experience and influencing predictions about the dog's behavior.

**Level 3: Thoughtseed - The Pullback Attractor and Agent**

- **"Dog" Thoughtseed:** The "dog" thoughtseed emerges as a dominant attractor state within the brain's dynamic landscape, integrating perceptual, conceptual, and emotional information from activated KDs and NPDs. It embodies a unified and meaningful concept of "dog," encompassing various attributes, behaviors, and associated experiences. As the thoughtseed evolves through interactions, it influences the cognitive processes that shape conscious awareness.
- **Agency and Cognitive Access:** The "dog" thoughtseed acts as an active agent, generating predictions about the dog's behavior, such as expecting it to continue playing fetch or approach the observer. These predictions are continuously compared to actual sensory input, leading to updates in the thoughtseed's internal model and the refinement of its representation in conscious awareness.
- **Attentional Modulation:** The thoughtseed exerts top-down control over attention, directing focus to specific aspects of the dog or its environment based on the observer's goals and interests. This modulation facilitates the selection of relevant sensory inputs for further processing in the global workspace.
- **Competition and Selection:** The "dog" thoughtseed competes with other potential thoughtseeds for dominance in cognitive processing. Its selection as the dominant thoughtseed is determined by its Expected Free Energy (EFE), considering its epistemic and pragmatic affordances. The thoughtseed's saliency, contextual relevance, and alignment with the living system's goals influence its likelihood of being selected. The emergence of the "dog" thoughtseed reflects its ability to provide the most accurate and parsimonious explanation for the current sensory input and the living system's goals, thereby minimizing free energy.
- **Action Selection:** The active "dog" thoughtseed guides the selection of actions toward the dog, such as petting it or calling its name. These actions are aimed at fulfilling the thoughtseed's predictions and minimizing free energy.

**Level 4: Meta-Cognitive Level**

- **Higher-Order Thoughtseeds:** Higher-order thoughtseeds, representing goals or intentions (e.g., "I want to play with the dog"), modulate the activity of the "dog" thoughtseed and influence its policy selection. These thoughtseeds exert top-down control over cognitive processes, shaping attention and the salience of different aspects of the "dog" thoughtseed. This allows for flexible and adaptive behavior, enabling the living system to prioritize certain aspects of the "dog" concept based on current goals and intentions.

- **Stream of Thoughts:** The continuous emergence, competition, and transition of thoughtseeds create a dynamic stream of thoughts within the global workspace. As the observer interacts with the dog and the environment, different thoughtseeds may activate, reflecting changes in attention and ongoing active inference. For example, the "dog" thoughtseed might transition to a "play" thoughtseed during a game of fetch or to a "fear" thoughtseed if the dog suddenly barks aggressively. This dynamic interplay of thoughtseeds, illustrates the living system's efforts to make sense of its environment, predict future events, and select actions that minimize surprise and achieve its goals.

## Code Link

# 10. Acknowledgments